\documentclass[conference]{IEEEtran}
\IEEEoverridecommandlockouts

\usepackage{amsmath}
\usepackage{amssymb}
\usepackage{graphicx}
\usepackage{booktabs}
\usepackage{threeparttable}
\usepackage{cite}
\usepackage{url}

\title{Implementation and Evaluation of BitNet Inference on a CGLA by Signed-Int4 Instructions}
\author{
\IEEEauthorblockN{Takuto Ando, Yasuhiko Nakashima}
\IEEEauthorblockA{Nara Institute of Science and Technology, Japan}
}

\begin{document}

\maketitle

\begin{abstract}
Large language model (LLM) inference transfers model weights and activations for every generated token, making memory traffic and its energy cost part of the decode path.
BitNet b1.58 represents its low-bit weights by ternary values and uses integer activations.
However, this arithmetic does not match conventional int8 or floating-point general matrix multiplication, and existing BitNet accelerators implement it in specialized datapaths.
We instead map this operation to a CPU-Grounded Linear Array (CGLA), a programmable ASIC with explicit direct memory access, local memories, and reusable compiler-visible integer lanes.
The mapping adds \texttt{OP\_SMA4} as a reusable signed-int4 multiply--accumulate instruction rather than a BitNet-only datapath.
Each ternary weight occupies one signed 4-bit lane.
Each int8 activation is split into two signed-int4 fragments and reconstructed by shift-and-add.
Frequency scaling of the 145\,MHz FPGA measurement to an 840\,MHz 28\,nm model gives a projected 0.390\,ns per signed-int4 product.
BitNet C++ decode with one offloaded call measures 2.52\,tokens/s, with most execution remaining on the host.
\end{abstract}


\section{Introduction}

Large language models (LLMs) are commonly served as autoregressive inference workloads on graphics processing units (GPUs), where every generated token requires another pass through model weights and activations.
These repeated transfers consume memory bandwidth and power throughout token generation.
As a result, decode latency and energy depend on memory bandwidth and memory-access energy as well as peak arithmetic throughput~\cite{ref:cgla_llm}.

Low-bit quantization addresses the traffic volume by reducing the amount of data moved through the memory hierarchy~\cite{ref:bitnet158,ref:bitnetcpp}.
BitNet b1.58 uses ternary weights and low-precision integer activations, but it still evaluates dot products between model weights and activations~\cite{ref:bitnet,ref:bitnet158,ref:bitnetcpp}.
The resulting arithmetic shape differs from conventional int8 or floating-point general matrix multiplication (GEMM).
Dedicated BitNet accelerators accommodate that shape by placing ternary arithmetic and data movement in specialized datapaths~\cite{ref:tellme,ref:tereffic,ref:vitallm,ref:bitrom,ref:lut158}.
Our objective is instead to map the ternary--int8 dot product to a low-power programmable accelerator without adding a BitNet-only datapath.

To retain a programmable execution path, we use a CPU-Grounded Linear Array (CGLA).
CGLA is a low-power programmable ASIC target with explicit DMA, local memory staging, and reusable compiler-visible integer lanes~\cite{ref:cgla_access,ref:cgla_llm}.
CGLA presents ordered logical rows to the compiler, so the same array and software stack can implement a low-bit mapping through the existing instruction interface.
The existing CGLA backend for llama.cpp, a C/C++ LLM inference runtime, provides \texttt{OP\_SMA8} for signed-int8 dot products in its Q8\_0 quantized format~\cite{ref:cgla_llm,ref:llamacpp}.
When directly applied to BitNet, the existing \texttt{OP\_SMA8} path allocates an 8-bit field to each ternary weight, although only three values are needed.

Therefore, we add \texttt{OP\_SMA4}, a reusable signed-int4 dot-product instruction.
A ternary weight $w\in\{-1,0,+1\}$ occupies one signed-int4 lane, while an int8 activation is divided into a signed low-nibble residue and an adjusted high fragment.
Two multiply--accumulate streams evaluate the fragments.
The host combines their partial sums by shift-and-add and corrects positive values from 120 to 127.
Although one \texttt{OP\_SMA4} stream carries eight packed signed-int4 products per 32-bit lane, the exact BitNet mapping requires two streams to cover an int8 activation.
The instruction accepts arbitrary signed-int4 operands, so its interface also supports non-ternary dot products.
This reuse matters when one accelerator serves several quantized kernels through the same compiler, memory interface, and runtime.
We integrate the mapping into the BitNet C++ runtime and separate three evaluation boundaries.
Fixed-word measurements isolate the cost of signed-int4 products.
Tiled projections include the second activation stream and repeated local memory module (LMM) tiles.
The integrated run adds the selected DMA/CGLA interval and the remaining host execution.

\begin{figure*}[t]
    \centering
    \includegraphics[width=0.98\textwidth]{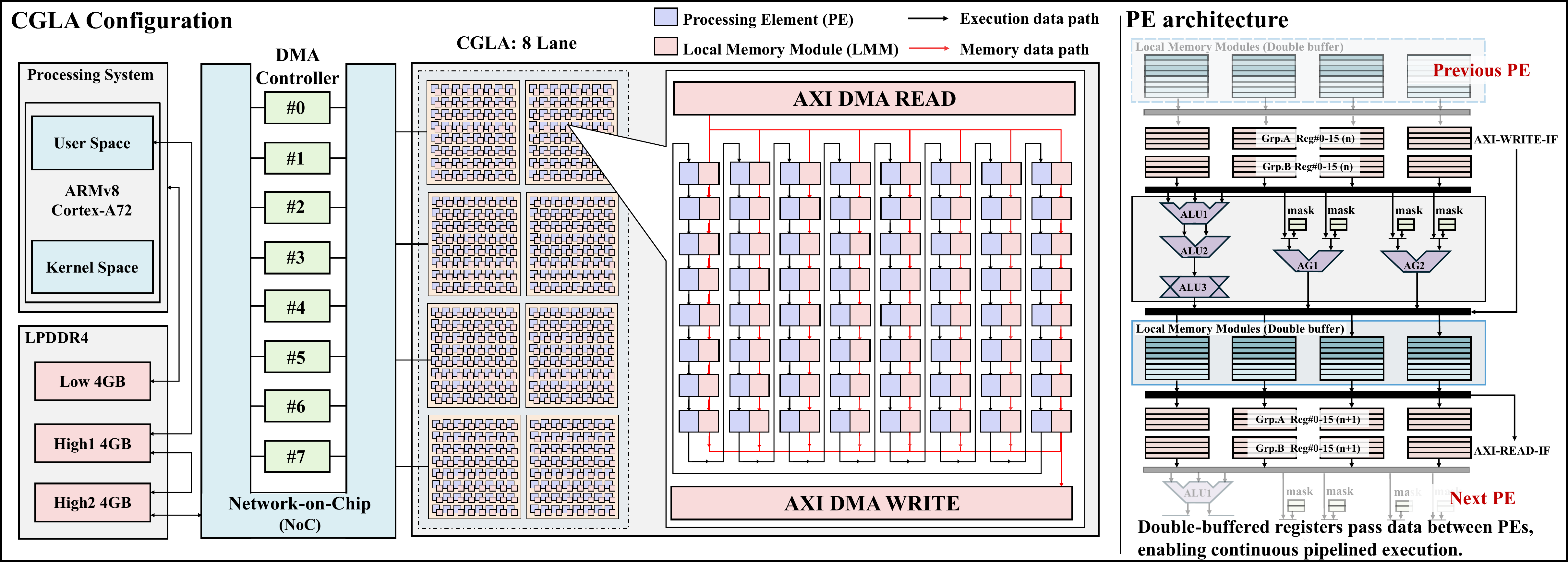}
    \caption{IMAX FPGA prototype and PE microarchitecture used in this work.
    The Processing System (PS) controls DMA transfers.
    The Programmable Logic (PL) contains the CGLA array.
    Each PE uses pipelined arithmetic logic units (ALUs), double-buffered registers, and local memories.}
    \label{fig:cgla-fpga-prototype}
\end{figure*}

The main contributions of this paper are as follows.
\begin{itemize}
\item We implement and evaluate \texttt{OP\_SMA4} as a reusable signed-int4 dot-product instruction and project its 2.26\,ns FPGA measurement to 0.390\,ns per product at 840\,MHz.
\item We validate the two-stream BitNet mapping over 4096 outputs using a sparse correction for values from 120 to 127.
\item We integrate the mapping into BitNet C++, where the 19-token decode measures 7.53\,s and 2.52\,tokens/s, including a 16.3\,ms offload interval.
\end{itemize}

The remainder of this paper is organized as follows.
Section~\ref{sec:related} reviews BitNet and programmable accelerator work.
Section~\ref{sec:method} describes the CGLA execution substrate and signed-int4 mapping, and Section~\ref{sec:evaluation} evaluates the instruction and integrated runtime boundary.
Section~\ref{sec:discussion} interprets the results and Section~\ref{sec:conclusion} concludes the paper.

\section{Related Work}
\label{sec:related}
\subsection{BitNet Models and Software Runtimes}

BitNet and BitNet b1.58 define the low-bit model family that includes the ternary-weight arithmetic implemented in this paper~\cite{ref:bitnet,ref:bitnet158}.
The BitNet b1.58 2B-4T technical report provides the model reference, and BitNet C++ provides the runtime tensor layout used in our implementation~\cite{ref:bitnet2b4t,ref:bitnetcpp}.
T-MAC and llama.cpp describe lookup, packing, and CPU-side low-bit execution around the arithmetic kernel~\cite{ref:tmac,ref:llamacpp}.
The evaluated BitLinear call combines packed I2\_S ternary-weight blocks with q8 activation blocks.
After the packed dot product, the runtime applies the activation and weight block scales to the integer partial sums.
CGLA offload retains the runtime tensor order, block arithmetic, and scaling sequence.
Model loading and token-generation control remain on the existing host path.

\begin{figure*}[!t]
\centering
\includegraphics[width=0.98\textwidth]{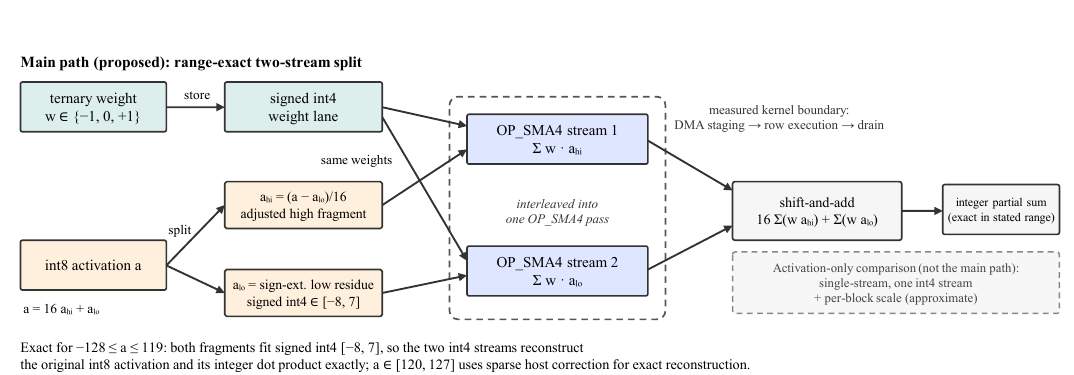}
\caption{Ternary--int8 dot-product mapping and CGLA timing components for \texttt{OP\_SMA4}.
A signed int4 lane stores the ternary weight, while the int8 activation is split into a signed low-nibble residue and an adjusted high fragment.
The two \texttt{OP\_SMA4} streams compute products for the adjusted high fragment and signed low residue, followed by shift-and-add reconstruction.}
\label{fig:sma4-mapping}
\end{figure*}

\subsection{Dedicated BitNet Hardware}

BitNet-oriented hardware papers primarily use dedicated datapaths.
TeLLMe and TerEffic target ternary LLM inference on FPGA platforms~\cite{ref:tellme,ref:tereffic}.
VitaLLM, BitROM, and LUT-based accelerator generation target ASIC, CiROM, or LUT-oriented designs~\cite{ref:vitallm,ref:bitrom,ref:lut158}.
These designs specialize the datapath for ternary-weight arithmetic.
Bit Fusion dynamically composes bit-level processing elements to match the operand widths of different DNN layers~\cite{ref:bitfusion}.
Our mapping instead uses a fixed signed-int4 instruction and software decomposition of int8 activations on the ordered CGLA array.
In contrast, our implementation uses a reusable signed-int4 instruction on a programmable spatial architecture.

\subsection{Coarse-Grained Reconfigurable Arrays (CGRAs) and CGLA for LLMs}

CGRA and CGLA systems have been evaluated for LLM and Transformer workloads.
PICACHU targets nonlinear LLM operations on a plug-in CGRA~\cite{ref:picachu}.
NX-CGRA supports representative Transformer kernels on a programmable CGRA~\cite{ref:nxcgra}.
Prior CGLA work evaluates LLM kernels and system-level transfer bottlenecks on a CGLA platform~\cite{ref:cgla_access,ref:cgla_llm}.
These studies do not map BitNet-style ternary BitLinear kernels onto CGRA/CGLA.
We map the ternary BitLinear kernel to a signed-int4 instruction and evaluate its instruction cost, tiled projection, and integrated-call boundary.

\subsection{CGLA Architecture and Compilation Model}

CGLA is a programmable ASIC architecture with a linear processing array and a software-controlled execution path~\cite{ref:cgla_access}.
CGLA reuses processing elements across kernels and presents its row-oriented linear array as ordered logical rows to the compiler.
The loop bodies are mapped without a two-dimensional placement-and-routing search, which shortens compilation~\cite{ref:cgla_access}.
CGLA exposes a fixed instruction interface rather than a per-kernel FPGA implementation flow, although row count, supported operations, local-memory capacity, and transfer bandwidth remain physical constraints.
Fig.~\ref{fig:cgla-fpga-prototype} shows the AMD Versal VPK180 prototype, where programmable logic hosts the CGLA core and a network-on-chip connects it to the processing system.
At 145\,MHz, PEs and local memory modules (LMMs) form a multilevel linear pipeline that stages operands near the processing elements.
Each PE combines pipelined arithmetic units with double-buffered register state, while LMMs hold the operands addressed by the generated row schedule.
Intermediate values follow the ordered execution path shown in Fig.~\ref{fig:cgla-fpga-prototype}, so compilation selects the row instructions and their data dependencies instead of two-dimensional PE coordinates.
During an offloaded call, the processing system places packed operands in DMA-visible DDR buffers and controls transfers between DDR and the LMMs.
The generated schedule advances the staged operands through ordered PE rows and drains the results to a host-visible buffer.
Kernel changes are expressed by the instruction sequence and operand layout at this boundary rather than by implementing a new FPGA bitstream for each operator.
Prior CGLA work uses this organization for sparse and dense kernels, fast Fourier transform (FFT)-like kernels, convolutional neural networks (CNNs), and LLM kernels~\cite{ref:cgla_access,ref:cgla_llm}.
For BitNet, we retain the DMA, LMM, synchronization, and row-scheduling path and add a reusable signed-int4 arithmetic instruction to the software-visible operation set.

\section{Proposed Method}
\label{sec:method}
\subsection{BitNet Ternary--int8 Dot Product}

BitNet b1.58 uses ternary weights $w\in\{-1,0,+1\}$, and the inner operation is the dot product $y = \sum_i w_i a_i$ with an 8-bit activation $a_i$~\cite{ref:bitnet158}.
\texttt{OP\_SMA4} packs eight signed-int4 products into one 32-bit lane, with each ternary weight stored directly and each int8 activation supplied as two fragments through the existing signed-lane interface.
The CGLA operation remains a packed signed multiply--accumulate, while BitNet-specific processing is confined to operand packing and post-drain reconstruction.
One original BitNet product $w_i a_i$ therefore corresponds to two signed-int4 fragment products in the exact path.
We use ``int4 product'' for one fragment product in the fixed-word instruction measurement and ``original MAC'' for one ternary--int8 product in the layer projection.
Exact layer projections count two fragment products for each original MAC.

\subsection{Host--CGLA Execution Sequence}

BitNet C++ retains model loading, tokenization, sampling, and layer scheduling on the host.
A selected ternary dot product enters the CGLA software stack at the packed-buffer boundary.
The host first prepares ternary weights and two activation fragments in DMA-visible buffers.
DMA stages one operand tile from the cacheable DDR path exposed by \texttt{/dev/mem} into the LMMs.
A generated CGLA schedule then executes the signed dot products over the PE rows.
After the 32-bit partial sums drain to a host-visible buffer, host code reconstructs and scales them before returning control to the BitNet C++ layer.
\texttt{OP\_SMA4} reuses the existing CGLA transfer and synchronization path and changes the packed operand format and the arithmetic instruction issued within the row schedule.
When one projection exceeds the LMM capacity, the software stack partitions it into tiles.
Each tile repeats operand staging, row execution, and drain, and its returned partial contributes to the same layer output.
Layer execution accumulates returned tile partials instead of retaining a full BitNet layer in the LMM.
Reported CGLA time includes DMA staging, row execution, and drain rather than only arithmetic latency.
Model loading, tokenization, sampling, and layer scheduling remain outside this interval.

\subsection{OP\_SMA4 Instruction and Mapping}
\label{sec:mapping}

We implement the BitNet dot product using the CGLA signed-int4 dot-product instruction.
For operand width $p$, each 32-bit lane contains $32/p$ signed fields, and one 64-bit instruction contains two lanes indexed by $\ell\in\{0,1\}$.
The instruction updates each 32-bit accumulator as follows.
\begin{equation}
d'_{\ell}=d_{\ell}+\sum_{j=0}^{32/p-1}\operatorname{sext}_p(x_{\ell,j})\operatorname{sext}_p(z_{\ell,j}),
\end{equation}
where $d_{\ell}$ is the previous accumulator value and $x_{\ell,j}$ and $z_{\ell,j}$ are the $j$th signed fields of lane $\ell$.
Here $\operatorname{sext}_p$ interprets a $p$-bit field as a signed two's-complement integer and sign-extends it to the accumulator width.
The proposed \texttt{OP\_SMA4} uses $p=4$, so each lane accumulates eight signed-int4 products and the two lanes are updated independently.

The BitNet path keeps each ternary weight in a signed int4 lane and represents each int8 activation with two signed int4 fragments, as shown in Fig.~\ref{fig:sma4-mapping}.

Each ternary weight $w$ is stored directly in a signed 4-bit lane.
We split each int8 activation as $a = 16a_{\mathrm{hi}} + a_{\mathrm{lo}}$.
Here $a_{\mathrm{lo}}$ is the sign-extended low-nibble residue.
The adjusted high fragment is $a_{\mathrm{hi}}=(a-a_{\mathrm{lo}})/16$.
Both fragments fit signed int4 only when $a_{\mathrm{hi}}$ and $a_{\mathrm{lo}}$ lie in $[-8,7]$.
For $-128\le a\le119$, both fragments fit in signed int4, so the two streams reconstruct the original int8 activation exactly.
Positive values from 120 to 127 make $a_{\mathrm{hi}}=8$, which is encoded as $-8$ in the signed-int4 stream.
Encoding $a_{\mathrm{hi}}=8$ as $-8$ lowers the reconstructed product by $256w$, so the host adds $256w$ for each affected element after receiving the CGLA partial sums.
Let $\tilde a_{\mathrm{hi}}$ denote the encoded signed high fragment, including $-8$ when the adjusted high value is 8.
The CGLA partial sums are $s_{\mathrm{hi}}=\sum_i w_i\tilde a_{\mathrm{hi},i}$ and $s_{\mathrm{lo}}=\sum_i w_i a_{\mathrm{lo},i}$.
Reconstruction produces a 32-bit integer partial dot, rather than an 8-bit output.
We compute the sparse correction and reconstructed integer dot product as follows.
\begin{equation}
c=256\sum_{i:a_i\in[120,127]}w_i,\qquad
y=16s_{\mathrm{hi}}+s_{\mathrm{lo}}+c.
\end{equation}
The correction is zero when the activation block contains no value from 120 to 127.
With correction $c$, the measured projection reconstructs the complete int8 range exactly.
The two activation streams pass through the same generated schedule and return separate 32-bit partial sums.
Host code reconstructs the integer dot product by shifting the high-fragment sum by four bits and adding the low-fragment sum.
For each q8 block, the reconstructed integer partial is multiplied by the activation scale, and the accumulated output is then multiplied by the I2\_S weight scale.

The following dataflow distinguishes arithmetic reconstruction from transfer and scaling.
\begin{center}
\small
\begin{tabular}{@{}p{0.15\columnwidth}p{0.76\columnwidth}@{}}
\toprule
Location & Operation \\
\midrule
Host & Pack weights, low fragments, encoded high fragments, and correction $c$. \\
CGLA & Accumulate $s_{\mathrm{lo}}$ and $s_{\mathrm{hi}}$ using two \texttt{OP\_SMA4} streams. \\
CGLA & In the fused-output variant, form $s_{\mathrm{lo}}+16s_{\mathrm{hi}}$ before output transfer. \\
Host & Add $c$, apply block scales, and accumulate the projection output. \\
\bottomrule
\end{tabular}
\end{center}
The unfused variant transfers the two sums separately and also performs their shift-and-add on the host.
CPYOUT denotes the runtime operation that copies the CGLA output buffer back to host-visible memory.
In the archived fused schedule, rows 0--1 calculate addresses and rows 2--3 issue the signed-int4 operations.
Row 5 reduces the lane halves, row 6 shifts the high sum, and row 7 adds the low sum.
Row 63 stores the combined partial, giving a 64-row span with unused intermediate rows.
Sparse correction and block scaling remain on the host after this store and output transfer.

\subsection{Activation-Only Int4 Comparison Baselines}
\label{sec:single-stream}

\begin{figure}[t]
\centering
\includegraphics[width=0.96\columnwidth]{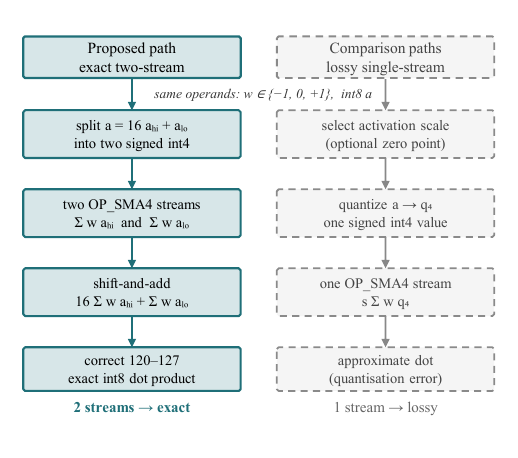}
\caption{Algorithmic paths compared in this work.
The proposed path splits each int8 activation into two signed-int4 fragments, evaluates two \texttt{OP\_SMA4} streams, and applies shift-and-add and sparse correction to reconstruct the ternary--int8 dot product exactly.
The comparison paths quantize the activation to one signed-int4 stream and produce approximate dot products.}
\label{fig:path-comparison}
\end{figure}

Fig.~\ref{fig:path-comparison} shows the algorithmic difference between the proposed exact two-stream mapping and the four activation-only single-stream int4 baselines.
We compare the proposed exact two-stream mapping with four activation-only single-stream int4 baselines.
The comparison measures path time and normalized partial-dot error after replacing two activation streams with one.
Exact reconstruction requires two activation streams because both signed-int4 fragments are required.
For comparison, we implement four single-stream approximations that quantize each int8 activation block to one signed-int4 block $q_4$ and use one \texttt{OP\_SMA4} stream.
The single-stream paths use the same ternary weights, signed-int4 instruction, and CGLA call boundary as the exact path.
Only the activation representation and host-side reconstruction differ, with one \texttt{OP\_SMA4} stream replacing the two exact streams and producing an approximate dot product.
Max-abs scaling uses $s=\max|a|/7$.
Power-of-two scaling rounds this scale to the nearest power of two.
Asymmetric affine scaling uses $s=(a_{\max}-a_{\min})/15$, $z=\operatorname{clip}_{[-8,7]}(-8-\operatorname{round}(a_{\min}/s))$, and $q_{4,i}=\operatorname{clip}_{[-8,7]}(\operatorname{round}(a_i/s)+z)$.
Affine reconstruction computes the dot product as $s(\sum_i w_iq_{4,i}-z\sum_i w_i)$.
Activation-MSE search evaluates ten scales from $0.5$ to $2.0$ times the max-abs scale and selects the scale with the smallest activation-reconstruction mean squared error.
For all four baselines, the quantization parameters are fixed before the CGLA call from the activation block alone.
Weights, CGLA outputs, and the exact ternary--int8 dot product are not used to select these parameters.
In affine reconstruction, $\sum_i w_i$ only removes the zero point after the CGLA result and does not select $s$ or $z$.

\section{Experiments and Discussion}
\label{sec:evaluation}

\subsection{Evaluation Setup}

The evaluation separates instruction timing, layer projection, and one integrated BitNet C++ call so that arithmetic cost is not conflated with runtime overhead.
Table~\ref{tab:device-platforms} summarizes the CGLA, Xeon, and RTX 4090 hardware specifications~\cite{ref:cgla_llm}.

\begin{table*}[t]
\centering
\begin{threeparttable}
\caption{Hardware specifications of the evaluated platforms.}
\label{tab:device-platforms}
\scriptsize
\begin{tabular*}{\textwidth}{@{\extracolsep{\fill}}llclclr@{}}
\toprule
\textbf{Device} & \textbf{Host} & \textbf{Compute units} & \textbf{Process} & \textbf{Frequency} & \textbf{Memory} & \textbf{Power [W]} \\
\midrule
CGLA 28\,nm model\tnote{c} & -- & 64 PEs\tnote{a} & 28\,nm & 840\,MHz & 32\,KiB LMM\tnote{b} & 1.82 \\
Intel Xeon W5-2465X & -- & 16 cores & Intel 7 & 3.10\,GHz & 512\,GB DDR5 & 65 \\
NVIDIA RTX 4090 & Xeon W5-2465X & 16384 CUDA cores & 5\,nm & 2520\,MHz & 24\,GB GDDR6X & 450 \\
\bottomrule
\end{tabular*}
\begin{tablenotes}[para,flushleft]
\scriptsize
\item[a] The CGLA compute-unit count denotes processing elements (PEs).
\item[b] The 32\,KiB LMM entry is the synthesis point covering the measured 30.1\,KiB resident window.
\item[c] Instruction and mapping timing is measured on the 145\,MHz FPGA prototype.
Only frequency-scaled timing and synthesis-power context use the 28\,nm CGLA row.
\end{tablenotes}
\end{threeparttable}
\end{table*}

Measurements use the CGLA software stack and the cacheable double-data-rate (DDR) path exposed by \texttt{/dev/mem}.
Reported kernel times are repeated-invocation wall-clock medians that include input staging, array execution, and output drain for each int4 loaded-word count.
Scalar host code checks the output.

The integrated experiment runs BitNet b1.58 2B-4T with two host threads and records ten 19-token runs as five pairs after model warm-up and prompt evaluation.

Tiled BitNet layer projections use the measured product costs.
For the device comparison, CGLA energy assigns 0.6485\,W to host execution and 2.07\,W to active offload.
CPU/GPU energy uses stated power or TDP-context estimates.

At the reported sizes, the int4 microbenchmark matches its scalar host reference.
The ternary$\times$int8 split path matches over 4096 outputs for tested inputs with $-128\le a\le119$.
In the selected projection comparison, positive values from 120 to 127 account for 2.07\% of the processed activation elements and use the sparse correction described in Section~\ref{sec:mapping}.
All CGLA partial sums match the packed-int4 host reference.
The corrected two-stream output has zero normalized integer-dot error.

\subsection{Signed-Int4 Instruction Results}

Each \texttt{OP\_SMA4} invocation includes input staging, array execution, and output drain.
Here $W$ denotes the number of loaded 32-bit operand words, each containing eight signed-int4 fields.
At $W=64$, $1024$, $4096$, $16384$, and $32768$ loaded 32-bit words, the kernel takes 15.0, 32.8, 87.5, 304, and 592\,$\mu$s.
A two-parameter fit over all five word counts gives a marginal cost of 0.0176\,$\mu$s per loaded word with $R^2>0.999$.
At $W=32768$, the measured cost is 2.26\,ns per signed-int4 product.
Scaling only the clock from 145 to 840\,MHz gives a projected cost of 0.390\,ns per product for the 28\,nm CGLA model.
At $W=32768$, each operand buffer occupies 128\,KiB in DDR, while each CGLA invocation uses a 30.1\,KiB resident LMM window.
The ASIC power row therefore uses the 32\,KiB LMM point.

\subsection{BitNet Projection and Integrated Measurement}

A BitNet projection layer with $m=6912$ and $k=2560$ contains $1.77\times10^7$ original ternary--int8 multiply--accumulate (MAC) operations and does not fit one LMM tile.
Multiplying the measured per-product cost by this operation count estimates 40\,ms for one fragment stream at 145\,MHz.
This is a layer-time estimate, not a directly timed invocation of that complete layer.
The exact \texttt{OP\_SMA4} mapping uses two fragment streams.
It projects to approximately 0.08\,s at 145\,MHz and 14\,ms at 840\,MHz.

Serial-row execution has a median elapsed time of 7.53\,s for 19 generated tokens, or 2.52\,tokens/s.
The five-run summary for this configuration reports 8.88\,ms of input staging, 5.82\,ms of execution, and 1.64\,ms of output drain.
These phases form the 16.3\,ms offload interval, leaving approximately 7.51\,s on the host.
The run enables one offloaded call after skipping 420 eligible calls, with two host threads and a $2560\times2560$ projection.
Its operand tiling uses eight output rows per batch and 640 physical words per row.
Output rows refer to matrix rows and must be distinguished from physical PE rows.
The 7.53\,s measurement is the runtime's accumulated 19-token decode-evaluation time after prompt evaluation.
It excludes model loading and prompt evaluation, so E2E below refers to this decode boundary.
The offload interval sums input staging, array execution, and output drain inside that boundary.
Embedding the CGLA call in the open-source BitNet C++ runtime preserves model loading, prompt evaluation, and token generation for direct E2E execution and measurement~\cite{ref:bitnetcpp}.
At the same packed-buffer boundary, the operand format, software-visible CGLA instruction, and generated row schedule can be changed without constructing a separate accelerator-specific inference runtime.
\begin{figure}[t]
\centering
\includegraphics[width=0.98\columnwidth]{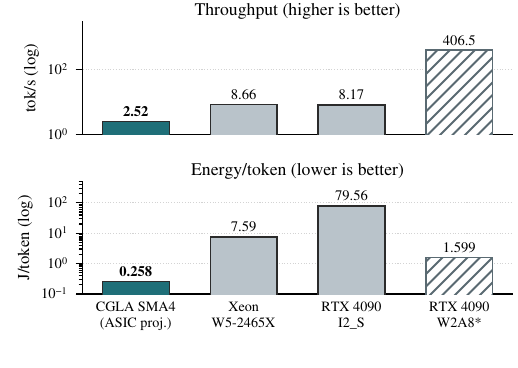}
\caption{BitNet b1.58 2B-4T decode throughput and estimated energy per token for the four evaluated implementations.
The CGLA row uses the fused five-run median and the 28\,nm CGLA projection.
I2\_S is the format-matched BitNet C++ CUDA path, whereas W2A8 is a PyTorch CUDA-graph reference.
Energy values use the row-specific power assumptions described in Section~\ref{sec:discussion}.}
\label{fig:device-comparison}
\end{figure}

Five pairs of 19-token runs compare the initial exact path with the eight-output-row fused-output exact path.
Fused-output execution reduces E2E time in all five pairs, with the median falling from 8.91 to 7.53\,s and active CGLA time from 551 to 16.3\,ms.
In a separate matched comparison, CGLA combines the two integer partials as $s_{\mathrm{lo}}+16s_{\mathrm{hi}}$ at the same row granularity.
This combination cuts CPYOUT volume from 1.56 to 0.78\,MB and the counter-reported transfer total from 2.07 to 1.29\,MB.
Median drain time decreases from 2.34 to 1.64\,ms, and the active interval decreases from 17.0 to 16.4\,ms in all five pairs.

Fig.~\ref{fig:device-comparison} reports a fused CGLA median of 2.52\,tokens/s and 0.258\,J/token over five runs.
For BitNet b1.58 2B-4T decode, the CPU/GPU rows span 8.17--406.5\,tokens/s and 1.599--79.56\,J/token.
Under the reported energy estimates, CGLA is the lowest-energy row among the four implementations.
In Fig.~\ref{fig:device-comparison}, W2A8 provides 406.5\,tokens/s compared with 2.52\,tokens/s for CGLA, whose modeled energy is 0.258\,J/token compared with 1.599\,J/token.
The preferred operating point therefore depends on the required decode rate and energy budget.
Because almost all decode time remains on the host, this comparison does not establish an accelerator-induced full-model speedup or energy saving.

\subsection{Activation-Only Baseline Comparison}
\label{sec:single-stream-results}
\begin{figure}[t]
\centering
\includegraphics[width=0.90\columnwidth]{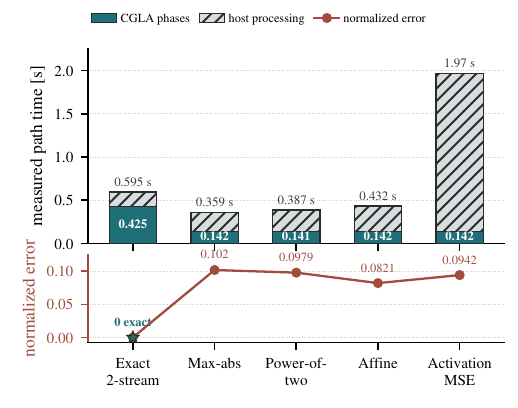}
\caption{Measured path-time decomposition and normalized partial-dot error for the proposed exact mapping and activation-only int4 comparison baselines on one selected projection.}
\label{fig:activation-baseline-comparison}
\end{figure}
We compare the proposed path and four activation-only baselines for one $2560\times2560$ projection with one decoded token and two host threads.
Fifteen performance runs from three prompts and five repetitions per prompt determine the path-time medians.
Three prompt-specific verification runs determine the error medians.
For integer partial dots $y_j$ and their approximations $\hat y_j$, normalized error is $\sum_j|\hat y_j-y_j|/\sum_j|y_j|$.
Each $j$ identifies an integer partial dot before runtime block scaling, rather than a generated token or an accuracy score.
Each baseline fixes its quantization parameters before the CGLA call.
The exact target is computed afterward only for reporting.
Exact two-stream execution takes 0.595\,s, including 0.425\,s in CGLA phases, and has zero sparse-corrected integer error.
Max-abs execution takes 0.359\,s, including 0.142\,s in CGLA phases, and gives a normalized error of 0.102.
Power-of-two scaling takes 0.387\,s and gives a normalized error of 0.0979.
Asymmetric affine execution takes 0.432\,s, including 0.142\,s in CGLA phases and host-side zero-point correction, and gives the lowest single-stream error of 0.0821.
Activation-MSE gives an error of 0.0942 and takes 1.97\,s because of host-side candidate evaluation.
Its CGLA phases still take 0.142\,s.
One-stream execution reduces CGLA time but introduces normalized partial-dot error, as shown in Fig.~\ref{fig:activation-baseline-comparison}.
All four baselines have CGLA phases because each executes its quantized activation stream through the same \texttt{OP\_SMA4} kernel.
No perplexity or downstream-task evaluation establishes whether these approximation errors are acceptable for deployment.
The exact two-stream path avoids this additional activation-quantization error at the integer-dot boundary.
Its local arithmetic check does not replace a full-model quality evaluation.

\subsection{Interpretation and Limitations}
\label{sec:discussion}

\texttt{OP\_SMA4} packs eight signed-int4 products per 32-bit lane and has a fitted marginal cost of 0.0176\,$\mu$s per loaded word.
Exact reconstruction evaluates two fragment products per original MAC.
Fused in-array combination halves CPYOUT volume and reduces drain time from 2.34 to 1.64\,ms.
Input staging and array execution remain on the active path, so the active interval decreases only from 17.0 to 16.4\,ms.
For one selected projection, the activation-only baselines trade exact reconstruction for shorter path time.
Affine quantization gives an error of 0.0821 at 27\% lower path time than the exact path.

The principal limitation is the evaluation boundary, which has three aspects.
First, fixed-word timing measures one int4 stream, and the tiled layer values are projections.
The integrated E2E run offloads one selected call rather than all BitNet layers.
Second, the 840\,MHz timing and 1.82\,W CGLA values use the existing 28\,nm model rather than silicon measurements.
Frequency-only scaling assumes that the measured interval scales with the core clock, including transfer overheads that may scale differently in silicon.
The instruction equation specifies the software-visible behavior, but the evaluation does not isolate the incremental decoder, register-path, or arithmetic-unit area and power.
Table~\ref{tab:device-platforms} lists the 450\,W device TDP of the RTX 4090.
Fig.~\ref{fig:device-comparison} retains the 650\,W combined host-plus-GPU TDP context used by the reported energy comparison.
Third, I2\_S is the format-matched BitNet C++ CUDA path, whereas W2A8 is a separate PyTorch CUDA-graph reference.
The four activation-only baselines are approximate paths excluded from the exactness result.
Further runtime integration will apply sparse correction across the offloaded layers and evaluate full-model execution with run-to-run timing variation.

\section{Conclusion}
\label{sec:conclusion}
In this paper, we mapped BitNet ternary--int8 dot products onto \texttt{OP\_SMA4}, a reusable signed-int4 instruction for a programmable CGLA.
At $W=32768$, the 145\,MHz FPGA measurement of 2.26\,ns per product corresponds to 0.390\,ns at 840\,MHz under the frequency-only 28\,nm CGLA projection.
The open-source BitNet C++ runtime measures 2.52\,tokens/s at the decode boundary with one offloaded call.
The CGLA call separately reports input staging, array execution, and output drain for further optimization.
Under the stated phase-accounted model, the fused CGLA path is estimated at 0.258\,J/token, below the three CPU/GPU rows in the device comparison.
Future work will optimize the measured host and DMA staging/drain intervals and extend the offload to full-model execution.

\bibliographystyle{IEEEtran}
\bibliography{references}

\end{document}